\documentclass[twocolumn,showpacs,preprintnumbers,amsmath,amssymb]{revtex4}
\pdfoutput=1

\usepackage{graphicx}
\usepackage{dcolumn}
\usepackage{bm}

\begin{document}

\preprint{NStaley}

\title{Suppression of conductance fluctuation in weakly disordered 
mesoscopic graphene samples near the charge neutral point}

\author{Neal E. Staley}
\author{Conor Puls}%
\author{Ying Liu}%
\email{liu@phys.psu.edu}
\affiliation{%
Department of Physics, The Pennsylvania State University, University Park, PA, 16802
}%

\date{\today}

\begin{abstract}
We measured the conductance fluctuation of bi- and trilayer graphene devices prepared on mechanical exfoliated graphene by an all-dry, lithography-free process using an ultrathin quartz filament as a shadow mask. Reproducible fluctuations in conductance as a function of applied gate voltage or magnetic field were found. As the gate voltage was tuned so that the graphene device was pushed to the charge neutral point, the amplitude of the conductance fluctuation was found to be suppressed quickly from a value consistent with universal conductance fluctuation when the devices were still well within weakly disordered regime. The possible physical origins of the suppression are discussed.  
\end{abstract}

                     
\maketitle

Novel effects in quantum interference of diffusive charge carriers (electrons or holes) in single layer graphene (1LG) - weak localization (WL) and weak antilocalization (WAL) - have been predicted \cite{AndoPRL02} and observed experimentally  \cite{GeimWL,DeHeerPRL07}. The honeycomb crystalline structure of 1LG features two equivalent sublattices, resulting in a band structure with a double valley degeneracy, a Berry phase of $\pi$, and WAL as well as WL in the weakly disordered regime, which corresponds to negative (WAL) and positive (WL) magnetoconductance (MC) \cite{AndoPRL02}. For short-range scatters such as vacancies that facilitate intervalley scattering, the valley degeneracy is lifted, leading to WL rather than WAL.  In mechanically exfoliated 1LG, positive MC characteristic of WL behavior was observed.  Its amplitude was found to be suppressed, attributed to the presence of a random field induced by a gradual structural deformation \cite{GeimWL,RothNature07}. In epitaxial graphene films grown on SiC with its electrical conductivity dominated by the first layer \cite{DeHeerPRL07}, however, MC was found to be positive at low and negative at high fields, corresponding to WL and WAL, respectively \cite{McCannWL}. In both cases, the results were puzzling - it is not clear why the long-length-scale structural deformation in mechanically exfoliated 1LG should have led to intervalley scattering needed for the observation of WL with a positive MC, and the presence of graphene layers adjacent to the first layer in epitaxial graphene did not destroy the valley degeneracy and WAL to yeild purely positive MC.

Another hallmark of the quantum interference effects of diffusive charge carriers is the universal conductance fluctuation (UCF) \cite{Altshuler85,LeeStone} observed in mesoscopic samples of conventional semiconductors and metals \cite{AltshulerBook}. The mesoscopic size of a weakly disordered sample results in the loss of self averaging of its physical properties. In the case of sweeping the gate voltage for semiconducting samples, the Fermi level moves, leading to different conductance. Ramping the magnetic field, on the other hand, varies the phase of the wave function of a charge carrier, which is well correlated within the dephasing length, $L_\phi$.  As the flux enclosed in the area defined by $L_\phi$ approaches a flux quantum, the phase is scrambled by 2$\pi$, again resulting in different conductance. The conductance fluctuation (CF) resulted from the lack of self averaging in a mesoscopic sample is on the order of $e^{2}/h$ independent of its shape and the level of disorder (within the weakly disordered regime), therefore referred to as UCF \cite{Altshuler85,LeeStone, AltshulerBook}. 

Different from WL/WAL, CF in graphene devices should not be affected by the subtle issues of the intra- $vs.$ intervalley scattering processes. Indeed, CF was seen in 1LG prepared by mechanical exfoliation \cite{GeimWL, KimNanoribbon} and epitaxial graphene films grown on SiC \cite{DeHeerScience06}. No suppression nor enhancement of the CF was identified in either system. However, we note that anomalously large CF with its amplitude larger than UCF at low disorder was predicted theoretically for 1LG, which is yet to be experimentally confirmed \cite{BeenakkerUCF}. We choose to focus on CF in bi- and trilayer graphene (2LG and 3LG). Theoretically no anomaly is expected for UCF in either system. Effects of WL in 2LG have been studied theoretically \cite{TrigonalWarp2LG} and experimentally \cite{WL2LG}. In Ref. \cite{WL2LG} CF was seen on top of WL MC but not quantified. For 3LG, however, neither WL/WAL nor CF has been studied. In this paper, we report observation of unexpected suppression of CF near the charge neutral point (CNP) in 2LG and 3LG that are well within the weakly disordered regime where UCF should have been seen.     

Our samples were fabricated using an all-dry, lithography-free technique. Heavily N-doped silicon with a 300-nm-thick thermally grown SiO$_{2}$ top layer was used as a substrate. A thin quartz filament was placed on a 2LG or 3LG flake and used as a shadow mask. Thicknesses of the flakes were determined by an optical method based on the color and faintness of the flakes, correlated also with atomic force microscopy (AFM) and Raman spectroscopy measurements. Two electrodes were formed by evaporation of Au. Two fine Au wires were then attached to each electrode by Ag epoxy, resulting in essentially two-wire devices.  The N-doped silicon was used as a back gate which produces 7 $\times$ $10^{10}$ carriers per volt. All devices measured in this study had the graphene connected to the positive polarity for the gate voltage bias, meaning that applying positive gate voltage corresponds to adding holes. All measurements were carried out in a cryostat with a base temperature of 1.2 K. The magnetic field was applied perpendicular to the graphene plane. The mobility of our devices was estimated for devices at high gate voltages using $\mu = (d/\epsilon _0 \epsilon)( \partial \sigma/\partial V_g)$, with its value ranging from 850 cm$^{2}$/Vs to 2400 cm$^{2}$/Vs at low temperatures. Details of our fabrication technique were reported previously \cite{NStaley}.

In Fig. 1a we show the conductance ($\it{G}$) $vs.$ gate voltage ($V_g$) on a 2LG sample with an area (between the two Au electrodes) of 0.25 $\mu$m long and 1 $\mu$m wide. In order to quantify the CF, a conductance background needs to be subtracted. Away from the CNP, we used a linear background in the $\it{G}$ $\it{vs.}$ $V_g$ for this purpose because $\sigma = \mu ne$ $\sim$$V_g$ is expected. A third-order polynomial was used near the CNP.  To account for any abrupt change in the slope of $G$ $\it{vs.}$ $V_g$, a piecewise background subtraction was employed. After background subtraction, reproducible CF as a function of gate voltage was clearly seen (Fig. 1b), suggesting that the CF is intrinsic.  A background is subtracted from the MC, $G$ $\it{vs.}$ $B$ data as well, to quantify the MC fluctuation (MCF). However, for MCF, the effects related to WL/WAL at the lowest fields, which are subjected to subtleties as discussed above, need to be treated with care. As shown in Fig. 2a, MC shows an overall conductance drop in the low-field limit. Similar features were found in all 2LG and 3LG devices we measured. While the low-field cusp in MC data is characteristic of WL effect \cite{LeeRamakrishnan}, we find that the MCF can become comparable in magnitude with that of the WL contribution near $B = 0$, making the subtraction of a MC background difficult. Given this, and that this work focuses on CF, we chose to subtract the MC background uniformly in all fields using a function of the form, $G_{BKGRD} = G_0+a\vert B \vert + bB^{2}$.  The linear term has its origin in a slight density gradient in the samples \cite{JainBook} and the quadratic term comes from the conventional MC. Essentially we ignore the WL/WAL effects expected at low fields, which fortunately does not affect the evaluation of MCF. As shown in Fig. 2b, data from two separate measurements again suggest that the MCF seen in the $G$ $\it{vs.}$ $B$ traces is intrinsic.

\begin{figure}
\includegraphics{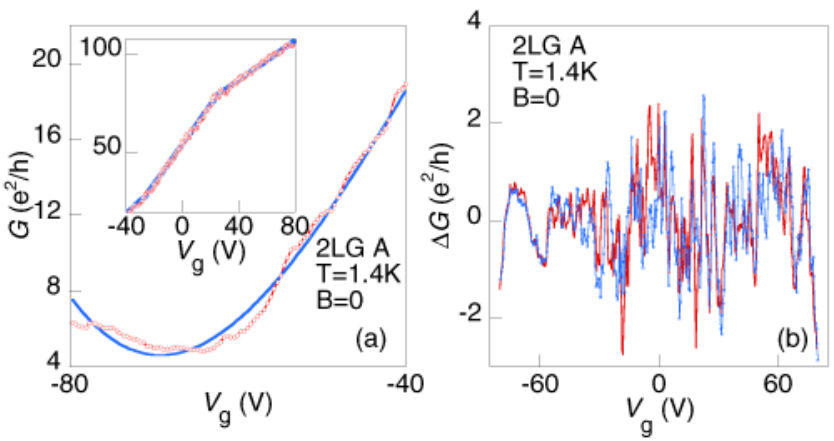}
\caption{(Color Online) (a) $\it{G}$ $\it{vs.}$ $V_g$ (red) for a 2LG sample showing CF near the charge neutral point (CNP).  Note that the CNP is shifted away from zero gate voltage due to doping from adsorbed gasses. Conductance background (blue) was shown.  Inset shows $\it{G}$ $\it{vs.}$ $V_g$ (red) and the background (blue) for the remaining range of gate voltages; (b)  $\Delta$$\it{G}$ $\it{vs.}$ $V_g$ for two back-to-back sweeps of gate voltage on the same sample. The conductance background shown in (a) was removed from the original conductance data.}
\end{figure}

\begin{figure}
\includegraphics{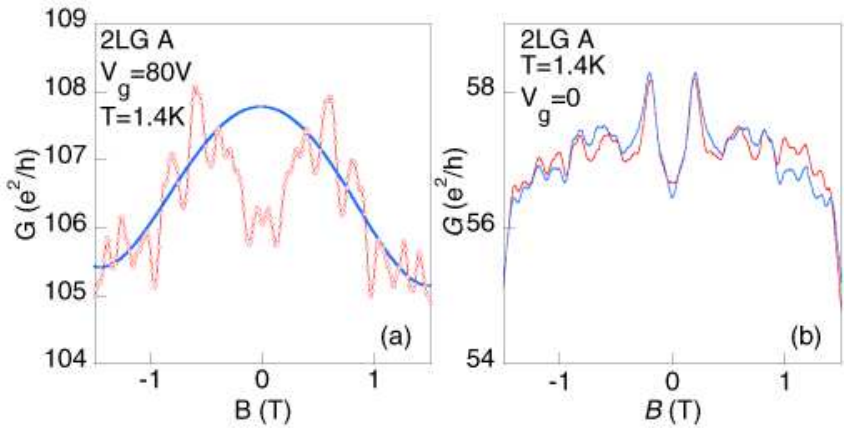}
\caption{(Color Online) (a) $\it{G}$ $\it{vs.}$ $B$ (Red) for the 2LG sample.   Blue line shows the background used for computing MCF;  (b) Back-to-back traces of $\it{G}$ $\it{vs.}$ $B$ for the 2LG sample. No background was subtracted for both traces in this plot.}
\end{figure}

\begin{figure}
\includegraphics{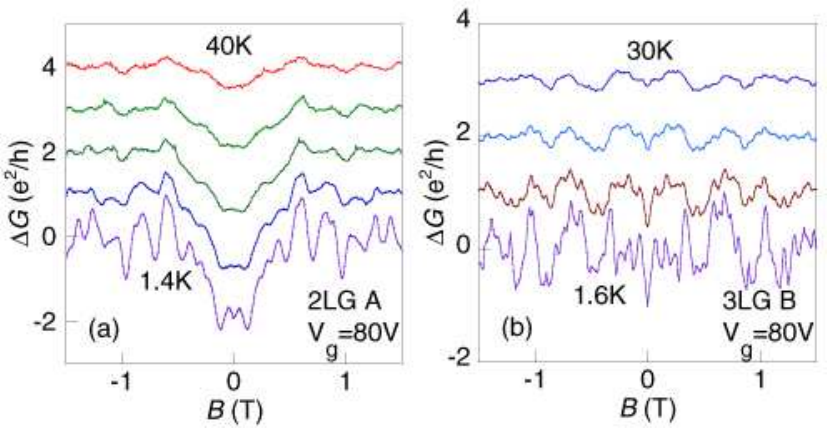}
\caption{(Color Online) (a) MCF in the 2LG sample at fixed temperature $T$ = 1.5, 10, 20, 30, and 40 K; (b) MCF in a 3LG sample at fixed temperature $T$ = 1.6, 10, 20, and 30 K; Traces in both plots except that for the lowest temperature were off set by adding a constant 1 $e^{2}$/h successively to the original MCF for clarity. }
\end{figure}

In Fig. 3 we show MCF traces at fixed temperatures for the 2LG  shown above and a 3LG sample, respectively. The MCF, with an amplitude of 1.5 $e^{2}/h$ peak to peak for the 2LG at $T$ = 1.4 K and 1.46 $e^{2}/h$ peak to peak for the 3LG at $T$ = 1.6 K, is robust, persisting to temperatures as high as 30 - 40 K. The characteristic intervals for MCF, related to the value of $L_\phi$, is seen in Fig. 3 to increase as the temperature increases, consistent with the picture that $L_\phi$ decreases with increasing temperature.  Even at 30-40K, the characteristic field intervals are seen to be on the order of 0.5 T, indicating that $L_\phi$ was on the order of 70 nm at these high temperatures, possible only for weakly disordered samples.  

In Fig. 4a, we show the $\it{G}$ $\it{vs.}$ $V_g$ for the 2LG sample.  The CNP for this device is seen at $V_g$ = -65 V with a two-dimensional (2D) conductivity of $\sigma _{2D}=1.2$ $e^{2}/h$ (Fig. 4a).  The amplitude of the CF as a function of $V_g$ was found to be roughly 2 $e^{2}$/h far from the CNP, decreasing monotonically to approximately 0.4 $e^{2}$/h as the CNP was approached (Fig. 4b). The MCF is suppressed from an amplitude of 1 $e^{2}$/h far from to 0.07 $e^{2}$/h near the CNP around the same gate voltage where the CF as a function of gate voltage was found to be suppressed. Similar behavior was found in a 3LG device with a CNP at $V_g$ = -80 V and a minimal conductivity $\sigma _{2D}=3.57$ $e^{2}/h$ (Fig. 4d). The value of CF as a function of $V_g$ was found to be reduced from roughly 0.7 to 0.1 $e^{2}$/h (Fig. 4e) as $V_g$ was varied.  The amplitude of MCF was found to decrease by an order of magnitude, from 0.15 $e^{2}$/h at $V_g$ = +80 V (somewhat smaller than typical UCF) to 0.02 $e^{2}$/h at $V_g$ = -80 V (near the CNP), occurring again at the same gate voltage when the CF was found to be suppressed.  While the close relation between the suppression in CF and MCF and the reduction in conductance may be consistent with the general idea that one can suppress CF by increasing disorder, the unexpected observation is that the suppression of CF (and MCF) started at $\sigma _{2D}=7.5$ $e^{2}$/h or a conductance of $\sim 30$ $e^{2}$/h, for the 2LG device (with an area of 0.25 $\mu$m long and 1 $\mu$m wide) and at $\sigma _{2D}=45$ $e^{2}$/h or a conductance of $\sim$30 $e^{2}$/h for the 3LG device (with an area of 3 $\mu$m long and 2 $\mu$m wide). In 2D, the conductivity is given by $\sigma _{2D}=(k_Fl)e^{2}$/$h$ where $k_F$ is the Fermi wave number and $l$ the elastic mean-free path. As a result, the corresponding $k_Fl$ values for which the CF/MCF started to be suppressed were 7.5 and 45, respectively, much larger than the strongly disordered limit of $k_Fl \sim 1$, thus placing both devices well within the weakly disordered regime.

\begin{figure}
\includegraphics{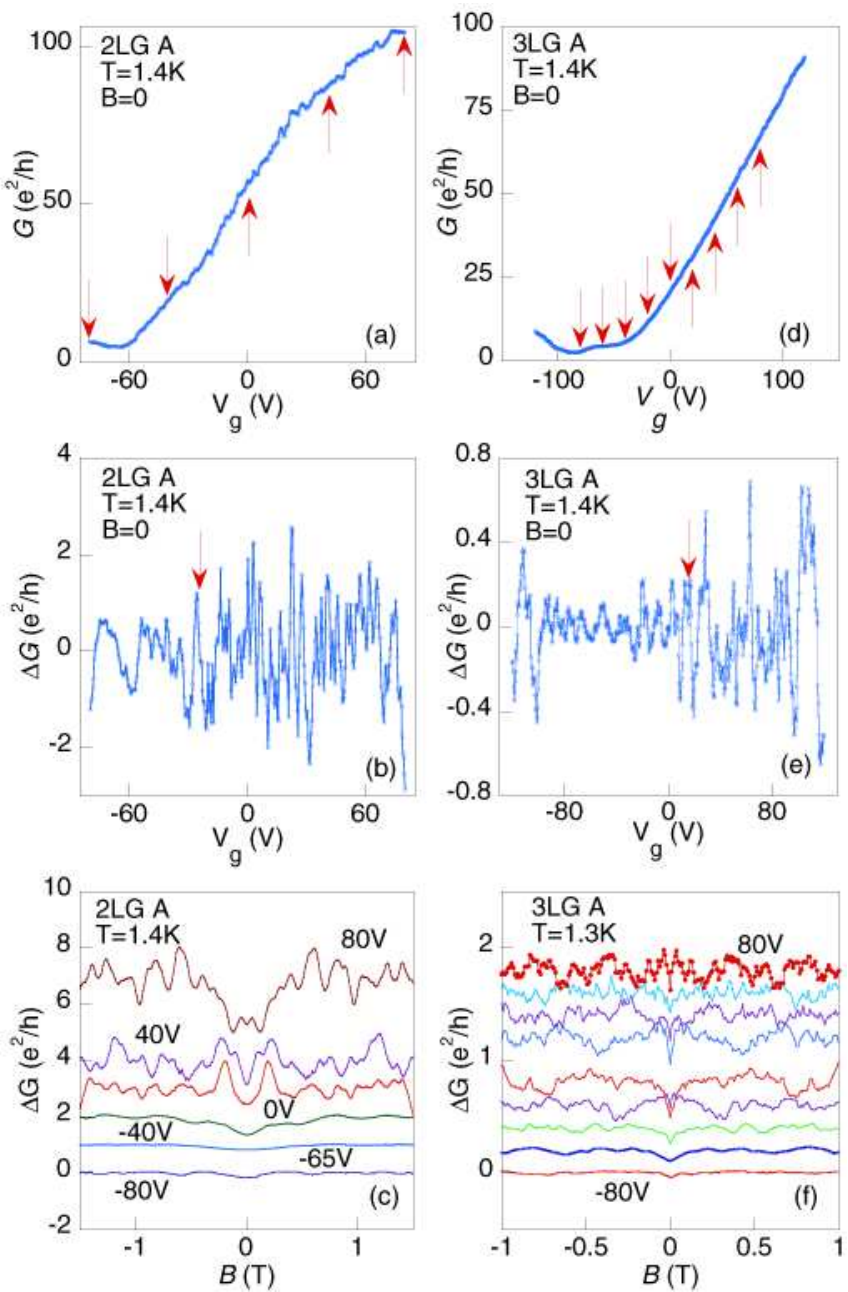}
\caption{(Color Online) (a) $\it{G}$ $\it{vs.}$ $V_g$ for the 2LG device. (b)  $\Delta$G $\it{vs.}$ $V_g$ for the same device; (c) $\Delta$G $\it{vs.}$ $B$ at several $V_g$ = -80, -65, 0, 40, and 80 V. (d) $G$ $\it{vs.}$ $V_g$ for a 3LG device. (e) $\Delta$ $\it{G}$ $\it{vs.}$ $V_g$ for the same sample;  (f) $\Delta$G $\it{vs.}$ $B$ at several $V_g$ = -80, -60, -40, -20, 0, 20, 40, 60, and 80 V. Traces in both (c) and (f) except that for $V_g$ = -80 V were offset by adding successively 1 $e^{2}$/h (c) or 0.2 $e^{2}$/h (f) to the original MCF for clarity. In (a) and (d) arrows indicate where $G$ $\it{vs.}$ $B$ was swept, and in (b) and (e) arrows indicate where CF started to be suppressed.}
\end{figure}

\begin{figure}
\includegraphics{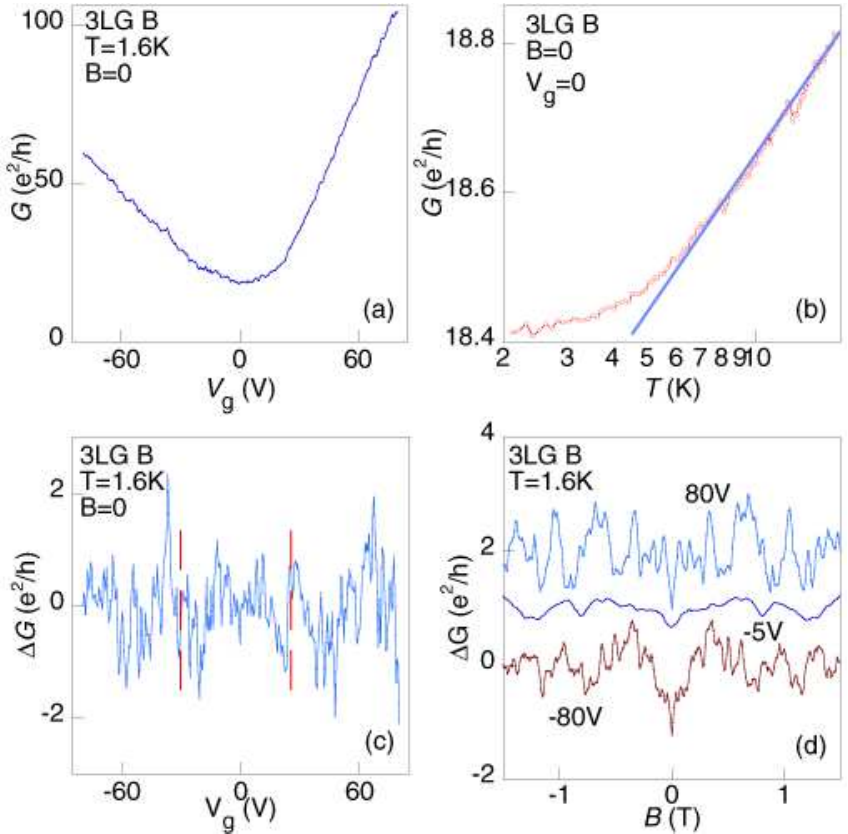}
\caption{\label{fig:epsart}  (Color Online) (a) $\it{G}$ $\it{vs.}$ $V_g$ for a 3LG device.  (b) $G$ $\it{vs.}$ $T$ near the CNP in semilogarithmic scale; (c) $\Delta$$\it{G}$ $\it{vs.}$ $V_g$ for the same sample, dashed lines indicate the $V_g$ values for which $G \approx 30$  $e^{2}$/h to help assess whether CF exhibits a suppression;  (d) $\Delta$$\it{G}$ $\it{vs.}$ $B$ at several $V_g$ = -80, -5, and 80 V. Traces except that for $V_g$ = -80 V were offset by adding successively 1 e$^{2}$/h to the original MCF for clarity. }
\end{figure}

The values of the conductance at which the CF/MCF started to be suppressed were found to be similar for both 2LG and 3LG devices shown above ($\sim$30 $e^{2}$/h), which raises an intriguing possibility of a ``critical'' conductance below which CF/MCF would be suppressed. In addition, a shift in CNP away from zero gate voltage was seen in both devices, most likely due to adsorbed gasses that function as charge donors/acceptors, which could be a cause of concern that the observed phenomena were resulted from the heavy chemical doping of the samples.  To address these issues we show below measurements on another 3LG sample (with a quadrilateral shape of 1 $\mu$m long, 1 $\mu$m wide at one electrode and 2 $\mu$m at the other).  The sample was minimally doped, with $\sigma _{2D}=12.7$ $e^{2}/h$ at the CNP (Fig. 5a), indicating that its overall disorder was smaller than that of the 2LG or the 3LG device shown in Fig. 4. The temperature dependence of the conductance was found to exhibit the logarithmic behavior (Fig. 5b) expected from WL theory, with the saturation of conductance drop at low temperatures attributable to the emergence of a cutoff length, the sample size, for the temperature-dependent inelastic scattering length, $l_i \sim T^{-p}$, where $\it{T}$ is the temperature and $\it{p}$ a positive constant. The amplitude of CF as a function of $V_g$ was found to vary little between 80 to -80 V with an amplitude of 1.3 $e^{2}$/h. Whether there is a suppression of CF when the conductance value fell below 30 $e^{2}$/h is not clear (Fig. 5c). The amplitude of MCF (Fig. 5d) was found to be suppressed from a value of 1 $e^{2}$/h to 0.2 $e^{2}$/h near the CNP ($k_Fl \approx 12.7$). Unfortunately no $\it{G}$ $vs.$ $B$ traces were taken at the intermediate gate voltages. As a result, the conductance value at which MCF started to be suppressed is not known for this 3LG device. Nevertheless, the suppression of MCF near the CNP is evident in this minimally doped 3LG sample, suggesting that the observed suppression of CF/MCF was not due to chemical doping.

Therefore UCF was indeed observed in our 2LG and 3LG samples away from the CNP. However, CF/MCF was found to be suppressed near the CNP.  We propose two possible scenarios that may account for the observed suppression of CF/MCF. First, the suppression could be due to the presence of edge states, of either the 2LG  \cite{NetoBilayer} or 3LG edges or 1LG edges \cite{NakadaFujitaPRB96} due to the uneven cut of the individual 2LG and 3LG flakes. In any case, edge states are likely to emerge near the Fermi energy of uncharged 2LG and 3LG. As the CNP is approached, the edge states may dominate the conductance of the sample as the density of states in the bulk is minimized. Since the edge states are highly one-dimensional, and quantum interferences among the edge states have to come from the opposite edges through the Au electrodes, the CF/MCF may be suppressed.  Alternatively, the suppression of the CF/MCF in our 2LG and 3LG devices could be caused by electronic inhomogeneities arising near the CNP. It was shown that nonuniform charge carrier density is present in 1LG near the CNP \cite{YacobyPuddles}, forming random ``puddles'' of electrons or holes.  While it has not been reported that similar puddles are also present in 2LG and 3LG, they seem to be quite plausible. It is not clear if charge carriers can maintain phase coherence across the boundary between the adjacent puddles, the presence of these puddles may affect CF as well as MCF.  More experimental and theoretical studies are needed before the suppression of CF/MCF observed in the present work can be fully understood.  

We would like to acknowledge useful discussions with P. Lammert, C. Berger, P. Campbell, J. Zhu, J. Jain, and A.H. Castro Neto. We would also like to thank T. Meyer and T. Mallouk for kindly allowing the use of their optical microscopes. This work is support in part by DOE under Grant No DE-FG02-04ER46159.

\bibliography{Bib.bib}

\end{document}